\documentclass[a4paper,11pt]{article}
\pdfoutput=1

\usepackage{jcappub}
\usepackage{grffile}

\usepackage{bm}

\newcommand{\avg}[1]{\langle #1 \rangle}
\newcommand{\G}[3]{U^{#1}_{#2, #3}}
\newcommand{\bourret}[1]{\overline{\overline{#1}} \,}
\newcommand{\tA}[1]{t_{#1}}
\newcommand{\tB}[1]{\tau_{#1}}
\newcommand{\dd}[0]{\ensuremath{\mathrm{d}}}%
\newcommand*{\ii}{\imath}
\newcommand{\Liouville}[0]{\mathcal{L}}
\newcommand{\dLiouville}[0]{\delta\mathcal{L}}
\renewcommand{\vec}[1]{\ensuremath{\bm{#1}}}%
\let\cdotOLD\cdot
\renewcommand{\cdot}[0]{\!\cdotOLD\!}%
\newcommand{\vhat}[1]{\ensuremath{\widehat{\bm{#1}}}}%
\newcommand{\gout}[0]{g_{\text{out}}}
\newcommand{\kout}[0]{k_{\text{out}}}
\newcommand{\kin}[0]{k_{\text{in}}}
\newcommand{\threej}[6]{\left( \begin{array}{ccc} #1 & #3 & #5 \\ #2 & #4 & #6 \end{array} \right)}
\newcommand{\sixj}[6]{\left\{ \begin{array}{ccc} #1 & #3 & #5 \\ #2 & #4 & #6 \end{array} \right\}}

\title{Cosmic Ray Small-Scale Anisotropies in Quasi-Linear Theory}
\date{\today}

\author[a,b,1]{P. Mertsch,\note{Corresponding author.}}
\author[b]{M. Ahlers}

\affiliation[a]{Institute for Theoretical Particle Physics and Cosmology (TTK), RWTH Aachen University,\\52056 Aachen, Germany}
\affiliation[b]{Niels Bohr International Academy, Niels Bohr Institute,\\Blegdamsvej 17, 2100 Copenhagen, Denmark}

\emailAdd{pmertsch@physik.rwth-aachen.de}
\emailAdd{markus.ahlers@nbi.ku.dk}

\abstract{
The distribution of arrival directions of cosmic rays is remarkably isotropic, which is a consequence of their repeated scattering in magnetic fields. Yet, high-statistics observatories like IceCube and HAWC have revealed the presence of small-scale structures at levels of 1 part in 10,000 at hundreds of TeV, which are not expected in typical diffusion models of cosmic rays. We follow up on the suggestion that these small-scale anisotropies are a result of cosmic ray streaming in a particular realisation of the turbulent magnetic field within a few scattering lengths in our local Galactic neighbourhood. So far, this hypothesis has been investigated mostly numerically, by tracking test particles through turbulent magnetic fields. For the first time, we present an analytical computation that through a perturbative approach allows predicting the angular power spectrum of cosmic ray arrival directions for a given model of turbulence. We illustrate this method for a simple, isotropic turbulence model and we find remarkable agreement with the results of numerical studies.
}

\begin{document}
\maketitle
\flushbottom

\section{Introduction}

The arrival directions of cosmic rays (CRs) are highly isotropic. Usually, this is explained as a consequence of pitch-angle scattering between CRs and turbulent magnetic fields. If the large-scale distribution of CR sources results in a spatial gradient, quasi-linear theory~\cite{1966ApJ...146..480J,1966PhFl....9.2377K,1967PhFl...10.2620H,1970ApJ...162.1049H,Jokipii1972} predicts a small dipole anisotropy. Yet, observations show fluctuations on smaller scales, down to $10^{\circ}$ degrees. These small-scale anisotropies are conveniently quantified by the angular-power spectrum of the relative intensity of cosmic rays or, equivalently, by the phase-space density $f(\vec{r}_\oplus,\vec{p},t)$ inferred by an observer at position $\vec{r}_\oplus$ and time $t$. In the following, we will study the power spectrum per unit square of the phase-space volume defined by
\begin{equation}\label{eq:Cl}
C_{\ell}(t) \equiv \frac{1}{4 \pi} \! \int \! \! \dd \vhat{p}_A \int \! \! \dd \vhat{p}_B P_{\ell} ( \vhat{p}_A \cdot \vhat{p}_B) f_A f_B\,,
\end{equation}
where we use the abbreviation $\vhat{p} = \vec{p} / |\vec{p}|$ and $f_A = f(\vec{r}_\oplus,\vec{p}_A,t)$, {\it etc.} Small-scale anisotropies are \emph{not} present in the usual quasi-linear theory with uniform pitch-angle scattering. (See however Ref.~\cite{Giacinti:2016tld}). For a recent review on observations and interpretations of the small-scale anisotropies see Ref.~\cite{Ahlers:2016rox} 

One of the arguably most attractive explanations of the small-scale anisotropies is that they are due to magnetic turbulence itself~\cite{Giacinti:2011mz,Ahlers:2013ima,Ahlers:2015dwa}. Standard quasi-linear theory only predicts the ensemble-averaged phase-space density $\avg{f}$ and we can therefore only predict the angular power spectrum $C^{\text{std}}_{\ell}$ obtained from Eq.~\eqref{eq:Cl} through $f_A f_B \to \avg{f_A}\avg{f_B}$. Under the commonly used assumptions $C^{\text{std}}_{\ell} \sim 0$ for $\ell \geq 2$. (See again Ref.~\cite{Giacinti:2016tld} for modifications to this simple picture.) However, it is easy to see that in the ensemble-average the angular power spectrum $\avg{ C_{\ell} }$ can have small-scale power, i.e.\ $\avg{ C_{\ell} } > C^{\text{std}}_{\ell}$, if $\avg{f_Af_B} > \avg{f_A}\avg{f_B}$. In other words, if there are correlations between the fluxes of CRs arriving under an angle \mbox{$\theta \equiv \arccos{( \vhat{p}_A \cdot \vhat{p}_B)} \sim \pi / \ell$} (with $\ell$ the orbital quantum number corresponding to this angle $\theta$) then the average angular power spectrum $\avg{ C_{\ell} }$, computed from the ensemble average of the product of phase-space densities, will be larger than the standard angular power spectrum $C^{\text{std}}_{\ell}$, computed from the product of ensemble-averaged phase-space densities. Therefore, correlations lead to small-scale anisotropies.

These correlations are to be expected if particles propagate through a turbulent magnetic field: Particles arriving under an angle $\theta$ will have experienced similar fields for a certain amount of time before observation. It can be motivated~\cite{Ahlers:2013ima} that this time is of the order $\tau_{\text{sc}} / (\ell (\ell + 1))$ where $\tau_{\text{sc}}$ is the scattering time. It is therefore ultimately the spatial correlations of the turbulent magnetic field that are reflected in the angular correlations of CR arrival directions. 

In the following we will predict the angular power spectrum Eq.~\eqref{eq:Cl} in an extended quasi-linear theory, taking into account the angular correlation between phase-space densities. We will consider the case with an isotropic turbulence tensor and without regular magnetic field. In this configuration, the unperturbed trajectories are straight lines, thus particles are propagating ballistically.

The remainder of this paper is organised as follows: In Sec.~\ref{sec:single}, we present a formalism describing the evolution in a random magnetic field of the ensemble-averaged cosmic ray phase-space density $\avg{f}$ from time $t_0$ to time $t$ by the propagator $\G{}{t}{t_0}$. We extend on this in Sec.~\ref{sec:pair} to treat the correlated evolution of the ensemble-average of a product of phase-space densities by a pair propagator. Evaluating the lowest order terms of the ensuing perturbative series, we formulate an ordinary differential equation for the ensemble-averaged angular power spectrum $\avg{C_{\ell}}$ and present an analytical expression for its steady-state. We fix the only free parameter of this model by a comparison with test particle simulations in Sec.~\ref{sec:validation}. In Sec.~\ref{sec:results}, we show the predicted angular power spectrum and compare to observations from HAWC and IceCube. We summarise and conclude in Sec.~\ref{sec:summary}.

\section{Single-Particle Propagator}
\label{sec:single}

In the following, we will make use of a diagrammatic formalism for solving stochastic differential equations, as used for instance in propagation of waves through random media. Here, we briefly review this formalism to fix our notation. We refer the interested reader to Refs.~\cite{Frisch:1968aa,1977JPlPh..18...49P} for details. For simplicity, we will assume relativistic cosmic rays and work in natural units, $c=1$.

The problem of propagation of (relativistic) charged particles through a static regular and turbulent magnetic field $\overline{\vec{B}}$ and $\vec{\delta B}(\vec{r})$ can be formulated using Liouville's equation for the phase-space density $f = f(\vec{r}, \vec{p}, t)$,
\begin{equation}
\partial_t f + \vhat{p} \cdot \vec{\nabla} f + \Liouville{} f = -\dLiouville{}(t) f\, ,
\label{eqn:Liouville}
\end{equation}
with the deterministic and stochastic Liouville operators
\begin{align}
\Liouville{} = - i \vec{\Omega} \cdot \vec{L} \quad \text{and} \quad \dLiouville{} = - i \vec{\omega}(\vec{r}) \cdot \vec{L} \, ,
\label{eqn:def_Liouville_operators}
\end{align}
where $\vec{\Omega} = q \overline{\vec{B}} / p_0$ and $\vec{\omega}(\vec{r}) = q \vec{\delta B}(\vec{r}) / p_0$ are the (relativistic) gyrovectors of the regular and turbulent field, respectively, and $L_i \equiv-i \epsilon_{ijk}p_j\partial_{p_k}$ are angular momentum operators.

In the following, we will assume that the spatial dependence of the phase-space density can be approximated by the first two terms of a Taylor expansion,
\begin{equation}
f(\vec{r}, \vec{p},t) \simeq {f_\oplus(\vec{p}, t)} + (\vec{r} - \vec{r}_\oplus) \cdot \nabla\overline{f} \, .
\label{eqn:gradient_ansatz}
\end{equation}
where $f_\oplus(\vec{p}, t) \equiv f(\vec{r}_\oplus,\vec{p}, t)$ denotes the local phase-space density and $\overline{f}$ is the local angular-averaged phase-space distribution. With this ansatz, the Liouville equation~\eqref{eqn:Liouville} evaluates to
\begin{equation}
\partial_t f_\oplus + \Liouville{} f_\oplus + \delta \mathcal{L}(t) f_\oplus \simeq - \vhat{p} \cdot \nabla\overline{f} \, .
\label{eqn:Liouville_with_gradient}
\end{equation}
Here, the stochastic Liouville operator depends on time, as the turbulent magnetic field is evaluated along the particle trajectory.

Eq.~\eqref{eqn:Liouville_with_gradient} can be formally solved as
\begin{equation}
f_\oplus(\vec{p}, t) \simeq \G{}{t}{t_0} f_\oplus(\vec{p}, t_0) - \int_{t_0}^t \dd t' \G{}{t}{t'} \vhat{p} \cdot \nabla\overline{f} = \G{}{t}{t_0} f_\oplus(\vec{p}, t_0) + \Delta \vec{r}(t_0)\cdot \nabla\overline{f} \, ,
\label{eqn:formal_solution}
\end{equation}
with $\Delta \vec{r}(t_0) \equiv \vec{r}(t_0)-\vec{r}_\oplus$ and the aid of the time-evolution operator (also called propagator), written using the time-ordered (``latest--to--left'') exponential,
\begin{align}
\G{}{t}{t_0} &= \mathcal{T} \exp \left[ - \int_{t_0}^t \dd t' \, (\Liouville{} + \dLiouville{} (t')) \right] = \G{(0)}{t}{t_0} \mathcal{T} \! \exp \left[ - \!\! \int_{t_0}^t \dd t' \left( \G{(0)}{t'}{t_0} \right)^{-1} \dLiouville{}(t') \G{(0)}{t'}{t_0} \right] .
\label{eqn:time_ordered_exponential}
\end{align}
Here, $\G{(0)}{t}{t_0}$ denotes the free propagator,
\begin{equation}
\G{(0)}{t}{t_0} = \exp \left[ - (t - t_0) \Liouville{} \right] \, .
\end{equation}

What complicates the solution of Eq.~\eqref{eqn:Liouville} is the stochastic nature of $\dLiouville(t')$. One can only hope to predict moments of the propagator, its first moment being the expectation value. In the Gaussian limit, the expectation value of the propagator, $\avg{ \G{}{t}{t_0} }$ contains only two-point functions of $\dLiouville(t')$, 
\begin{equation}
\avg{ \dLiouville{}(t_n) \dLiouville{}(t_{n-1}) \mathellipsis \dLiouville{}(t_1) } = \avg{ \dLiouville{}(t_n) \dLiouville{}(t_{n-1}) } \mathellipsis \avg{ \dLiouville{}(t_1) \dLiouville{}(t_0) } + \text{permut.} \, ,
\end{equation}
for even $n$ and vanishes identically for odd $n$. The expansion of Eq.~\eqref{eqn:time_ordered_exponential} becomes algebraically complex very quickly. It can be diagrammatically written in a more economic form,
\begin{equation}
\vcenter{\hbox{
\includegraphics[scale=1,trim={0.5cm 0.6cm 0.5cm 0.7cm}, clip=false]{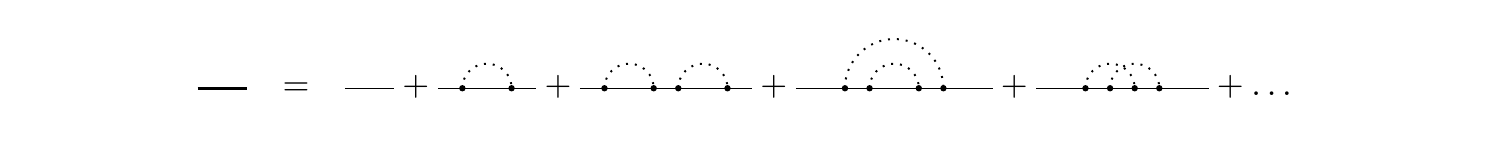}
}}
\end{equation}
Here, solid lines represent free single-particle propagators $\G{(0)}{t''}{t'}$, dots correspond to insertions of $\dLiouville{}(t')$ and dotted lines connecting such dots represent the expectation value of the two $\dLiouville$'s that it connects. All intermediate time variables are integrated over.

All connected diagrams can be resummed into the so-called mass operator,
\begin{equation}
\vcenter{\hbox{
\includegraphics[scale=1,trim={0.7cm 0.5cm 0.8cm 0.7cm}, clip=false]{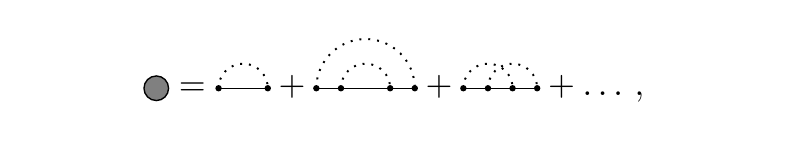}
}}
\end{equation}
such that the series for the propagator takes the simple form
\begin{equation}
\vcenter{\hbox{
\includegraphics[scale=1,trim={0.2cm 0.7cm 0.1cm 0.8cm}, clip=false]{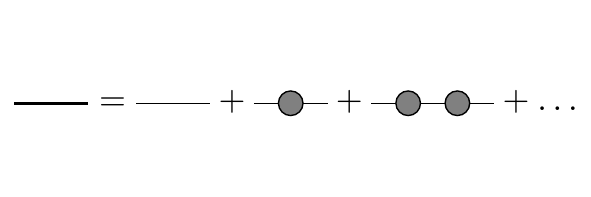}
}}
\end{equation}
The mass operator is difficult to evaluate at all orders, but approximating it with its lowest order term results in the so-called Bourret approximation to the single particle propagator,
\begin{equation}
\vcenter{\hbox{
\includegraphics[scale=1,trim={0 0.8cm 0 0.7cm}, clip=false]{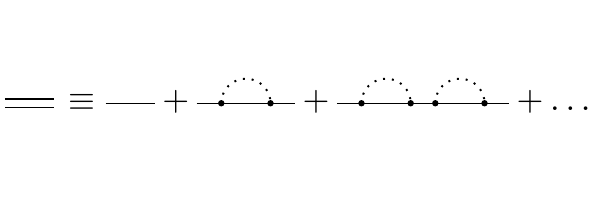}
}}
\label{eqn:Bourret_single1}
\end{equation}
This series can now be resummed. In the simple case of vanishing regular magnetic field ($\vec{\Omega} = \vec{0}$) the unperturbed trajectories are just straight lines and one finds~\cite{Casse:2001be}
\begin{equation}
\avg{ \G{}{t}{t_0} } \simeq \bourret{ \G{}{t}{t_0} }\equiv \G{(0)}{t}{t_0} e^{- (t-t_0) \nu\vec{L}^2/
2} \, .
\label{eqn:Bourret_single2}
\end{equation}
The parameter $\nu$ contains integrals over the two-point functions of the turbulent field $\omega(\vec{r})$.

\section{Pair Propagator}
\label{sec:pair}

The small-scale anisotropies are a consequence of the fact that the trajectories of a pair of CRs are correlated for a (finite) amount of time before observation. Therefore, we need to consider the ensemble average of products of phase-space densities when computing the angular power spectrum. Note that in standard quasi-linear theory, we compute the ensemble average of single phase-space densities and are therefore missing the correlations between pairs of CR particles.

In the following, we will use the abbreviations $f_A(t) \equiv f_\oplus(\vec{p}_A,t)$, {\it etc}. From Eq.~\eqref{eqn:formal_solution}, we find for the ensemble average of the product of phase-space densities,
\begin{align}
\avg{f_A(t) f_B^*(t)} 
\simeq & \avg{ \G{A}{t}{t_0} \G{B*}{t}{t_0} } \avg{ f_A(t_0) f_B^*(t_0) } + \avg{ (\Delta \vec{r}_A(t_0)\cdot\nabla\overline{f}) \G{B*}{t}{t_0} } \avg{ f^*_B(t_0) } 
\nonumber
\\ & + \avg{ (\Delta \vec{r}_B(t_0)\cdot\nabla\overline{f}^*) \G{A}{t}{t_0} } \avg{ f_A(t_0) } + \avg{ (\Delta \vec{r}_A(t_0)\cdot\nabla\overline{f})(\Delta \vec{r}_B(t_0)\cdot\nabla\overline{f}^*) }\,.
\label{eqn:formal_solution_product}
\end{align}
In our previous analysis~\cite{Ahlers:2015dwa} we identified the last term on the right-hand-side of Eq.~\eqref{eqn:formal_solution_product} as the term that determines the asymptotic behavior of the power-spectrum for large look-back times, $t-t_0\gg\nu^{-1}$. Note that we have assumed that correlations between the propagators and the initial state $f_\oplus(\vec{p},t_0)$ can be ignored. In the following, we will take a different approach and aim to establish a differential equation for the angular power spectrum in quasi-linear theory based on Eq.~\eqref{eqn:formal_solution_product}. This differential equation will describe the temporal evolution of the angular power spectrum \emph{locally}, that is at one position and we will thus consider the infinitesimal limit $\Delta T \to 0$.

By virtue of the Bethe-Salpeter equation~\cite{1951PhRv...84.1232S}, the double propagator can be expanded into a perturbative series. This series has a diagrammatic representation, somewhat similar to Feynman diagrams employed in quantum field theory,
\begin{equation}
\vcenter{\hbox{
\includegraphics[scale=0.98,trim={0.6cm 0.4cm 0.2cm 0.6cm}, clip=false]{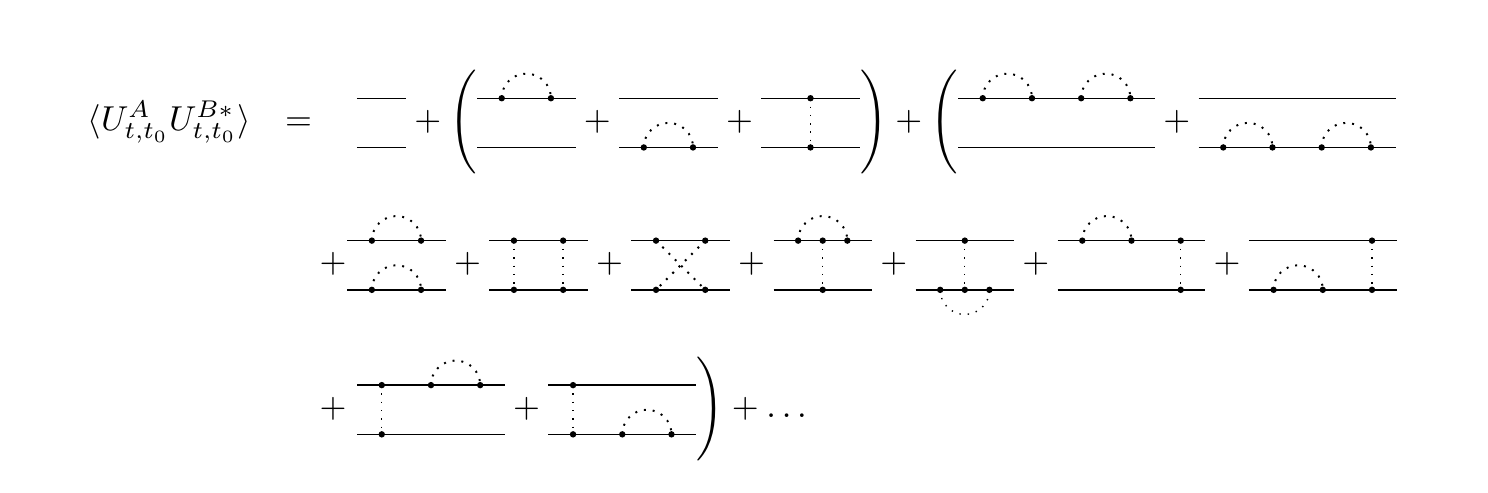}
}}
\label{eqn:diagrams}
\end{equation}
If the dashed lines are connecting the Liouville operators $\dLiouville{}$ of two different particles A and B, then this can be considered an interaction between particles A and B mediated by the correlation structure of the turbulent magnetic field. It is the repeated action of these ``interactions'' that is inducing the correlations between particles A and B.

While Eq.~\eqref{eqn:formal_solution_product} allows computing the angular power spectrum anytime after preparing the initial state, $f(\vhat{p}, t_0)$, evaluating or even resumming all diagrams of Eq.~\eqref{eqn:diagrams} in all generality seems challenging at the very least. Instead, we seek to approximate the identity~(\ref{eqn:formal_solution_product}) by the stationary solution of a  differential equation with respect to a small step in look-back time $\Delta T \equiv t-t_0$,
\begin{align}
\frac{1 - \avg{ \G{A}{t}{t_0} \G{B*}{t}{t_0} }}{\Delta T} \avg{ f_A(t_0)f^*_{B}(t_0) }
&\simeq (\overline{f} - 3 \vhat{p}_A \cdot {\bf K} \cdot \nabla\overline{f}) \left(\frac{\Delta\vec{r}_B}{\Delta T} \cdot \nabla\overline{f}\right) \nonumber
\\ & + \left(\frac{\Delta\vec{r}_A}{\Delta T} \cdot \nabla\overline{f}\right) (\overline{f} - 3 \vhat{p}_B \cdot {\bf K} \cdot \nabla\overline{f}) + \mathcal{O}(\Delta T)\,,
\end{align}
where we applied the quasi-stationary solution of the diffusion equation $\avg{ f_A(t_0) } \simeq \overline{f}-3 \vhat{p}_A \cdot {\bf K} \cdot \nabla\overline{f}$.
In this limit $\Delta T \to 0$, we can approximate $\Delta\vec{r}/\Delta T \simeq - \vhat{p}$. This allows writing down an ordinary differential equation for the angular power spectrum $C_{\ell}$,
\begin{equation}
A_{\ell \ell_0} C_{\ell_0}(t) \simeq \frac{8 \pi}{9} K \left| \nabla\overline{f} \right|^2 \delta_{\ell 1} \, ,
\label{eqn:ODE}
\end{equation}
where we assume isotropic diffusion $K_{ij} = K\delta_{ij}$ and define the transition matrix
\begin{equation}
A_{\ell \ell_0}(t) 
= \lim_{t_0 \to t} \frac{\delta_{\ell \ell_0} - M_{\ell \ell_0}(t, t_0)}{t - t_0}\,,
\label{eqn:A}
\end{equation}
and where
\begin{align}
M_{\ell \ell_0}(t, {t_0}) & = \frac{1}{4 \pi} \int \dd \vhat{p}_A \int \dd \vhat{p}_B P_{\ell} (\vhat{p}_A \cdot \vhat{p}_B) \avg{ \G{A}{t}{t_0} \G{B*}{t}{t_0} } \frac{2 \ell_0 + 1}{4 \pi} P_{\ell_0} (\vhat{p}_A \cdot \vhat{p}_B) \, . \label{eqn:mixing_matrix}
\end{align}
Once we have computed $M_{\ell \ell_0}(t, {t_0})$, it is easy to find the steady-state angular power spectrum $C^{\text{stdy}}$ by solving
\begin{equation}
A_{\ell \ell_0} C^{\text{stdy}}_{\ell_0}(t) = \frac{8 \pi}{9} K \left| \nabla\overline{f} \right|^2 \delta_{\ell 1} \, .
\label{eqn:ODE_stdy}
\end{equation}

In evaluating $\avg{ \G{A}{t}{t_0} \G{B*}{t}{t_0} }$, we confine ourselves to considering the leading and next-to-leading order terms, that is the first line of Eq.~\eqref{eqn:diagrams}. We label the contributions of those diagrams to the double propagator as follows,
\begin{align}
&\vcenter{\hbox{
\includegraphics[scale=1,trim={0.5cm 0cm 1cm 0.5cm}, clip=false]{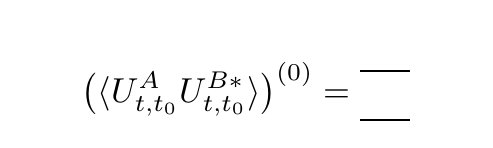}
}} \, ,
\vcenter{\hbox{
\includegraphics[scale=1,trim={0.5cm 0.3cm 0.7cm 0.7cm}, clip=false]{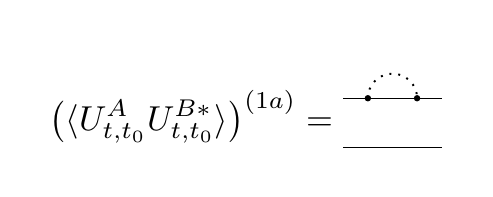}
}} \, ,
\\
&\vcenter{\hbox{
\includegraphics[scale=1,trim={0.5cm 0.1cm 0.7cm 0.5cm}, clip=false]{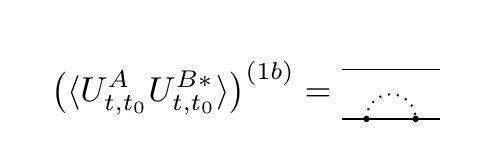}
}} \, ,
\vcenter{\hbox{
\includegraphics[scale=1,trim={0.5cm 0cm 0.5cm 0.5cm}, clip=false]{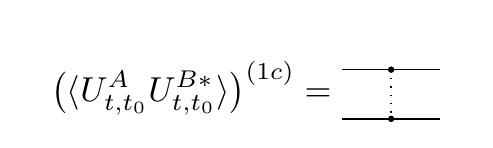}
}} \, .
\end{align}
The leading order term, the free double propagator,
\begin{equation}
\left( \avg{ \G{A}{t}{t_0} \G{B*}{t}{t_0} } \right)^{(0)} = 1 \, ,
\end{equation}
is trivial since the free single propagator is trivial, \mbox{$\G{(0)}{t}{t_0} = 1$}, in the limit of vanishing regular magnetic field ($\vec{\Omega} = \vec{0}$). The contribution of $\left( \avg{ \G{A}{t}{t_0} \G{B*}{t}{t_0} } \right)^{(0)}$ to $M_{\ell \ell_0}(t, {t_0})$ thus takes the simple form
\begin{align}
M_{\ell \ell_0}^{(0)} =& \delta_{\ell \ell_0} \, . \label{eqn:M0}
\end{align}

The first next-to-leading order term (1a) can be computed assuming ballistic trajectories, \mbox{$\vec{r}(t') = \vec{r}_\oplus - (t - t')\vhat{p}$} and an isotropic turbulence tensor,
\begin{align}
\avg{ \tilde{\omega}_i(\vec{k}) \tilde{\omega}^*_j(\vec{k}') } = \frac{g(k)}{k^2} \left( \delta_{ij} - \widehat{k}_i \widehat{k}_j \right) \delta(\vec{k} - \vec{k}') \, .
\end{align}
With the help of a plane wave expansion we find
\begin{align}
\left( \avg{ \G{A}{t}{t_0} \G{B*}{t}{t_0} } \right)^{(1a)} 
&= \int_{t_0}^t \dd \tA{2} \int_{t_0}^{t_2} \dd \tA{1} \G{A(0)}{t}{\tA{2}} \avg{ \delta\mathcal{L}^A_{\tA{2}} \G{A(0)}{\tA{2}}{\tA{1}} \delta\mathcal{L}^A_{\tA{1}} } \G{A(0)}{\tA{1}}{t_0} \G{B*(0)}{t}{t_0} \nonumber
\\ &= -\sum_{\ell_A} (2 \ell_A + 1) \ii^{\ell_A} \Lambda_{\ell_A}(t-t_0) \int \dd \widehat{k} P_{\ell_A}( \vhat{k} \cdot \vhat{p}_A ) \left( \delta_{ij} - \widehat{k}_i \widehat{k}_j \right) L_i^A L_j^A\,,
\label{eqn:double1a}
\end{align}
where $P_{\ell}(\cdotOLD)$ denotes the Legendre polynomial of degree $\ell$ and we introduce the quantity
\begin{align}
\Lambda_{\ell_A}(\Delta T) &= \int_{0}^{\Delta T} \dd T \int_{0}^{T} \dd \tau \int \dd k \, g(k) j_{\ell_A}(k \tau) \, ,
\end{align}
where $ j_{\ell}(\cdotOLD)$ is the spherical Bessel function of the first kind.

Computing the contribution of $\left( \avg{ \G{A}{t}{t_0} \G{B*}{t}{t_0} } \right)^{(1a)} $ to the mixing matrix $M_{\ell \ell_0}(t, {t_0})$ via Eq.~\eqref{eqn:mixing_matrix} we find
\begin{align}
M_{\ell \ell_0}^{(1a)} =& - \frac{8 \pi}{3} \delta_{\ell \ell_0} \! \left( \! \Lambda_0(\Delta T) - \frac{1}{2} \Lambda_2(\Delta T) \! \right) \! \ell (\ell+1) \, .
\label{eqn:M1a}
\end{align}

There is evidence that the energy spectral density of interstellar turbulent magnetic fields follows a power law in wavenumber, with theoretically motivated values of the spectral index of $-7/2$ or $-11/3$~\cite{Elmegreen:2004wj}. Here, we will adopt a band-limited white noise spectrum, that is $g(k) = \gout{}$ if $\kout{} \leq k < \kin{}$ and $0$ otherwise. Below we will show that for this spectrum, the limit $t_0 \to t$ of Eq.~\eqref{eqn:A} only exists if we simultaneously let $\kin{} \to \infty$, while keeping $\alpha \equiv \kin{} \Delta T$ finite. Physically this means that as we let $\Delta T \to 0$, we need to also extend the turbulence spectrum to arbitrarily small scales such that particles can experience changes in the magnetic field during the time $\Delta T$. The parameter $\alpha$ encodes how many wavelengths of the smallest modes the particle traverses in the time interval $\Delta T$. We consider this to be a free parameter and determine it by comparing with numerical simulations and observational data below. We then find
\begin{align}
M_{\ell \ell_0}^{(1a)} = - \pi \frac{\gout{}}{\kout{}} r \delta_{\ell \ell_0} \Big\{ & - \frac{1}{\alpha} \cos \alpha + 2 \alpha \, {}_2 F_3\left( \frac{1}{2}, \frac{1}{2}; \frac{3}{2}, \frac{3}{2}, \frac{3}{2}; - \left( \frac{\alpha}{2} \right)^2 \right) \nonumber
\\ & + \alpha^2 \sin \alpha - \text{Si} [\alpha] \Big\} \ell (\ell+1) \, . \label{eqn:M1a_limit}
\end{align}
For $\kout{} \Delta T \ll 1$, $M_{\ell \ell_0}^{(1a)} / \Delta T$ is a function of $\alpha$ only. Specifically, 
\begin{equation}
\frac{M_{\ell \ell_0}^{(1a)}}{\Delta T} \propto \left\{
\begin{array}{l l}
\alpha & \text{for } \alpha \ll 1 \, , \\
\ln \alpha & \text{for } 1 \ll \alpha \ll \kin{}/\kout{} \, , \\
\text{const.} & \text{for } \alpha \gg \kin{}/\kout{} \, .
\end{array}
\right.
\end{equation}
These cases correspond, respectively, to the particle travelling a distance less than $1/\kin{}$, between $1/\kin{}$ and $1/\kout{}$ and more than $1/\kout{}$ in the time $\Delta T$.

The first interacting contribution (1c) is
\begin{equation}
\int_{t_0}^t \dd \tA{1} \int_{t_0}^t \dd \tB{1} \G{(0)A}{t}{\tA{1}} \G{(0)B*}{t}{\tB{1}} \avg{ \delta\mathcal{L}^A_{\tA{1}} \delta\mathcal{L}^{B*}_{\tB{1}} } \G{(0)A}{\tA{1}}{t_0} \G{(0)B*}{\tB{1}}{t_0} \, .
\end{equation}
We use Eq.~\eqref{eqn:def_Liouville_operators}, Fourier transform the correlation function and, again assuming ballistic trajectories, \mbox{${\bf r}_A(t_0) = {\bf r}_\oplus - (t-t_0)\vhat{p}_A$}, perform a free wave expansion for the exponential factors $\exp[\ii \vec{k} \cdot \vec{r}(t) ]$. This leads to
\begin{align}
& \int_{t_0}^t \dd \tA{1} \int_{t_0}^t \dd \tB{1} \G{(0)A}{t}{\tA{1}} \G{(0)B*}{t}{\tB{1}} \avg{ \delta\mathcal{L}^A_{\tA{1}} \delta\mathcal{L}^{B*}_{\tB{1}} } \G{(0)A}{\tA{1}}{t_0} \G{(0)B*}{\tB{1}}{t_0} \nonumber
\\ & = (4 \pi)^2 \int \dd \widehat{k} \int \dd k \, \avg{ \tilde{\omega}_i(\vec{k}) \tilde{\omega}^*_j(\vec{k}') } \sum_{\substack{\ell_A,m_A\\\ell_B,m_B}} \ii^{-\ell_A + \ell_B} \int_0^{k \Delta T} \dd \tA{1}' j_{\ell_A}(\tA{1}') \int_0^{k \Delta T} \dd \tB{1}' j_{\ell_B}(\tB{1}') \nonumber
\\ & \times Y_{\ell_A m_A}(\vhat{k}) Y^*_{\ell_B m_B}(\vhat{k}) Y^*_{\ell_A m_A}(\vhat{p}_A)Y_{\ell_B m_B}(\vhat{p}_B) L^A_i L^{B*}_j \, .
\end{align}

In the following, we split $(\delta_{ij} - \widehat{k}_i \widehat{k}_j)$ into a monopole and a quadrupole contribution (in $\vhat{k}$),
\begin{equation}
\left( \delta_{ij} - \widehat{k}_i \widehat{k}_j \right) = \frac{2}{3} \delta_{ij} + \left( \frac{1}{3} \delta_{ij} - \widehat{k}_i \widehat{k}_j \right) \, .
\end{equation}
leading to $M_{\ell \ell_0}^{(1c,0)}$ and $M_{\ell \ell_0}^{(1c,2)}$, respectively. We find
\begin{align}
M_{\ell \ell_0}^{(1c,0)} &= \frac{8 \pi}{3} \sum_{\ell_A} (2 \ell_A + 1) \kappa_{\ell_A \ell_A}(t-t_0) \threej{\ell}{0}{\ell_A}{0}{\ell_0}{0}^2 (2 \ell_0 + 1) \ell_0 (\ell_0 + 1) \, ,
\label{eqn:M1c0}
\end{align}
where $( \cdot )$ denotes the Wigner $3j$-symbol and with the triple integral
\begin{align}
\kappa_{\ell_A \ell_B} (\Delta T) &\equiv \int_{\kout{}}^{\kin{}} \mathrm{d} k \frac{g(k)}{k^2} h_{\ell_A} \left( k, \Delta T \right) h_{\ell_B} \left( k, \Delta T \right) \nonumber
\\ &= \frac{\gout{}}{\kout{}} \int\displaylimits_{1}^{\kin{}/\kout{}} \mathrm{d} k' \, k'^{-2} \!\!\!\! \int\displaylimits_0^{k' \kout{} \Delta T} \!\!\!\!\mathrm{d} \tA{1}' j_{\ell_A}(\tA{1}') \!\!\!\! \int\displaylimits_0^{k' \kout{} \Delta T} \!\!\!\! \mathrm{d} \tB{1}' j_{\ell_B}(\tB{1}') \, .
\end{align}

The quadrupole contribution requires significantly more algebra, but eventually reads
\begin{align}
M_{\ell,\ell_0}^{(1c,2)} & = \frac{4 \pi}{3} \frac{\ell_0 (\ell_0+1) (2 \ell_0 + 1)}{\threej{\ell_0}{0}{2}{0}{\ell_0}{0}} (-1)^{\ell_0} \sum_{\ell_A,\ell_B} \ii^{\ell_A + \ell_B} \kappa_{\ell_A \ell_B} (t-t_0) (2 \ell_A + 1)(2 \ell_B + 1) \nonumber
\\ & \times 
\sixj{2}{\ell}{\ell_0}{\ell_A}{\ell_0}{\ell_B} \threej{2}{0}{\ell_A}{0}{\ell_B}{0} \threej{\ell}{0}{\ell_A}{0}{\ell_0}{0} \threej{\ell}{0}{\ell_B}{0}{\ell_0}{0}\,. \label{eqn:M1c2}
\end{align}
The curly brackets in Eq.~\eqref{eqn:M1c2} denote the Wigner 6$j$-symbol. Also $M_{\ell \ell_0}^{(1c)} / \Delta T$ shows some simple dependence on $\alpha$, 
\begin{equation}
\frac{M_{\ell \ell_0}^{(1c)}}{\Delta T} \propto \left\{
\begin{array}{l l}
\alpha^{2 + 2 |\ell - \ell_0|} & \text{for } \alpha \ll 1 \, , \\
\alpha & \text{for } 1 \ll \alpha \ll \kin{}/\kout{} \, .
\end{array}
\right.
\end{equation}

Eventually, we compute $M_{\ell \ell_0} = M_{\ell \ell_0}^{(0)} + 2 M_{\ell \ell_0}^{(1a)} + M_{\ell \ell_0}^{(1c,0)} + M_{\ell \ell_0}^{(1c,2)}$ (see Eqs.~(\ref{eqn:M0}), (\ref{eqn:M1a_limit}), (\ref{eqn:M1c0}) and (\ref{eqn:M1c2})), determine $A_{\ell \ell_0}$ from Eq.~\eqref{eqn:A} and find the steady-state angular power spectrum from Eq.~\eqref{eqn:ODE_stdy}.

\section{Validation}
\label{sec:validation}

In order to validate the results of our analytical computation we now compare to numerical results following the method in Ref.~\cite{Ahlers:2015dwa}. The power spectrum can be derived from the last term of Eq.~\eqref{eqn:formal_solution_product} in the limit $\Delta T \to \infty$. We have back-tracked test particles in isotropic turbulent magnetic fields with band-limited white-noise spectrum. We have not assumed any regular component. Specifically, we have chosen $\kout{} r_g = 10^{-3}$ and $\kin{} r_g = 10^2$, with $r_g$ the particles gyroradius in the RMS turbulent field. This guarantees that there is a broad enough range of wavenumbers to be in resonance with ($r_g k_{\text{res}} \approx 1$, $k_{\text{res}}$ being the resonant wavenumber) while satisfying the requirement $\kin{}/\kout{} \gg 1$. The numerical backtracking results in a set of trajectories that converge at $\vec{r}_\oplus$. Thanks to Liouville's theorem, we can use this to compute the angular distribution at position $\vec{r}_\oplus$ and time $t$ by assuming a certain phase-space density at time $t_0$. To make the connection with our analytical approach, we adopt the same gradient dependence as in Eq.~\eqref{eqn:gradient_ansatz}. For each of 100 random realisations of the turbulent magnetic field, we compute the angular power spectrum from the phase-space density at position $\vec{r}_\oplus$ and time $t$ and finally compute the ensemble averaged angular power spectrum.

\begin{figure}[t]
\centering
\includegraphics[scale=1]{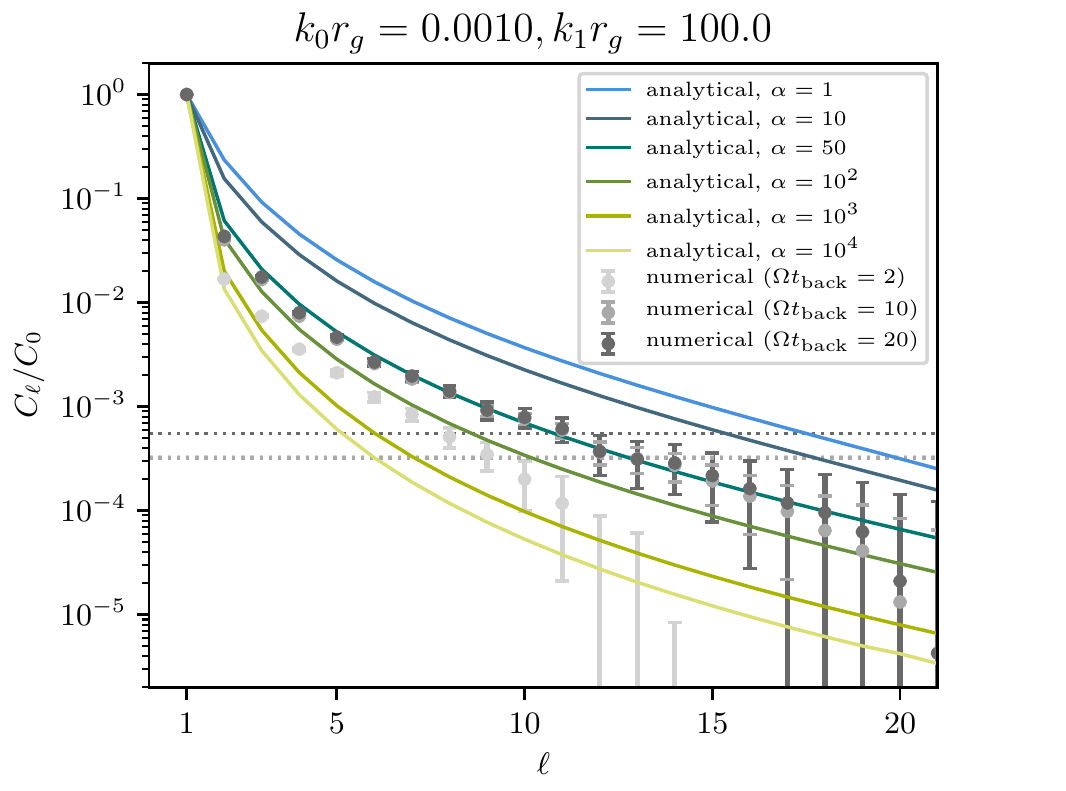}
\caption{Validation of the analytical method by comparison with numerical result. The data points show the angular power spectrum determined in test particle simulations for three different backtracking times $t_{\text{back}}$ after subtraction of the estimated shot noise. The shot noise levels due to the finite number of trajectories is indicated by the horizontal dashed lines. For comparison, the lines show the results of our analytical approach with different values of the free parameter $\alpha = \kin{} \Delta T$. Fixing this free parameter to $\kin{} \Delta T \approx 50$ results in excellent agreement with the test particle simulations.}
\label{fig1}
\end{figure}

In Fig.~\ref{fig1}, we show this ensemble averaged angular power spectrum for three different backtracking times $t_{\text{back}}$. It can be seen that the angular power spectrum converges to an asymptotic form for $\Omega t_{\text{back}} \gtrsim 10$ where $\Omega$ is the gyro frequency in the RMS turbulent field. (See also Ref.~\cite{Ahlers:2015dwa}.) Comparing with the analytical results allows fixing the free parameter, $\alpha = \kin{} \Delta T$, for which we otherwise only have the constraint $\alpha \gg 1$. It appears that $\alpha \approx 50$ gives excellent agreement between analytical and numerical results.

\section{Results}
\label{sec:results}

\begin{figure}[t]
\centering
\includegraphics[scale=1]{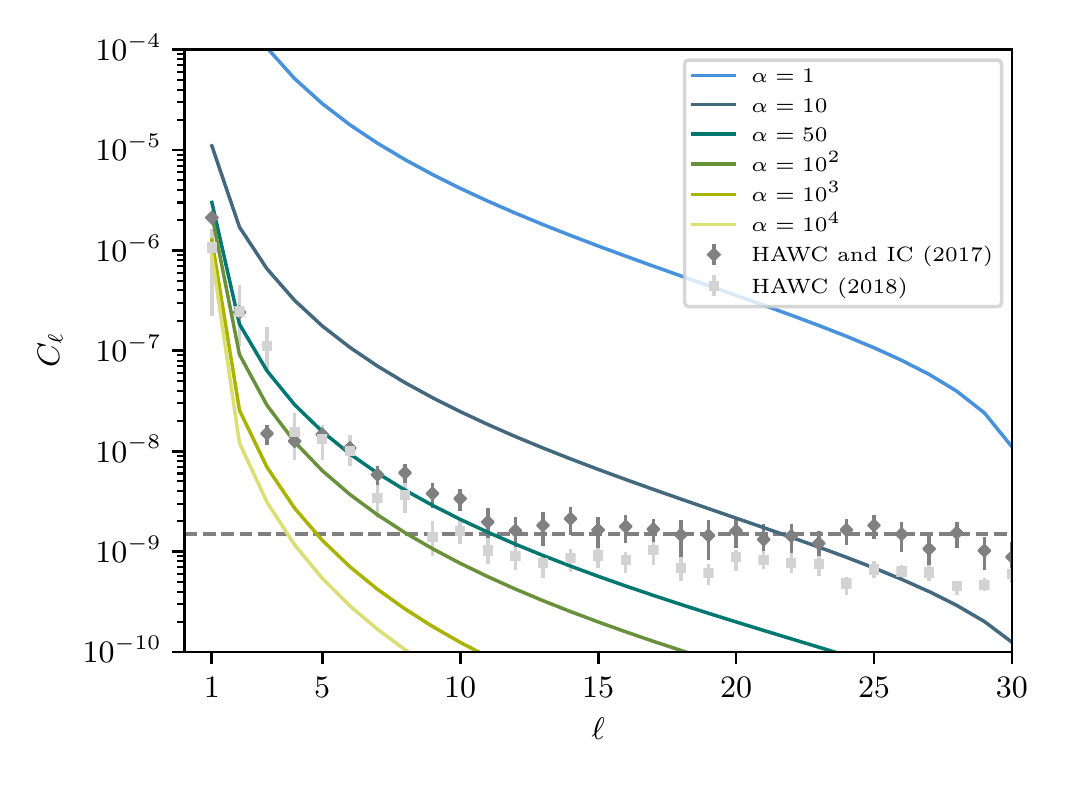}\\[-0.5cm]
\includegraphics[scale=1]{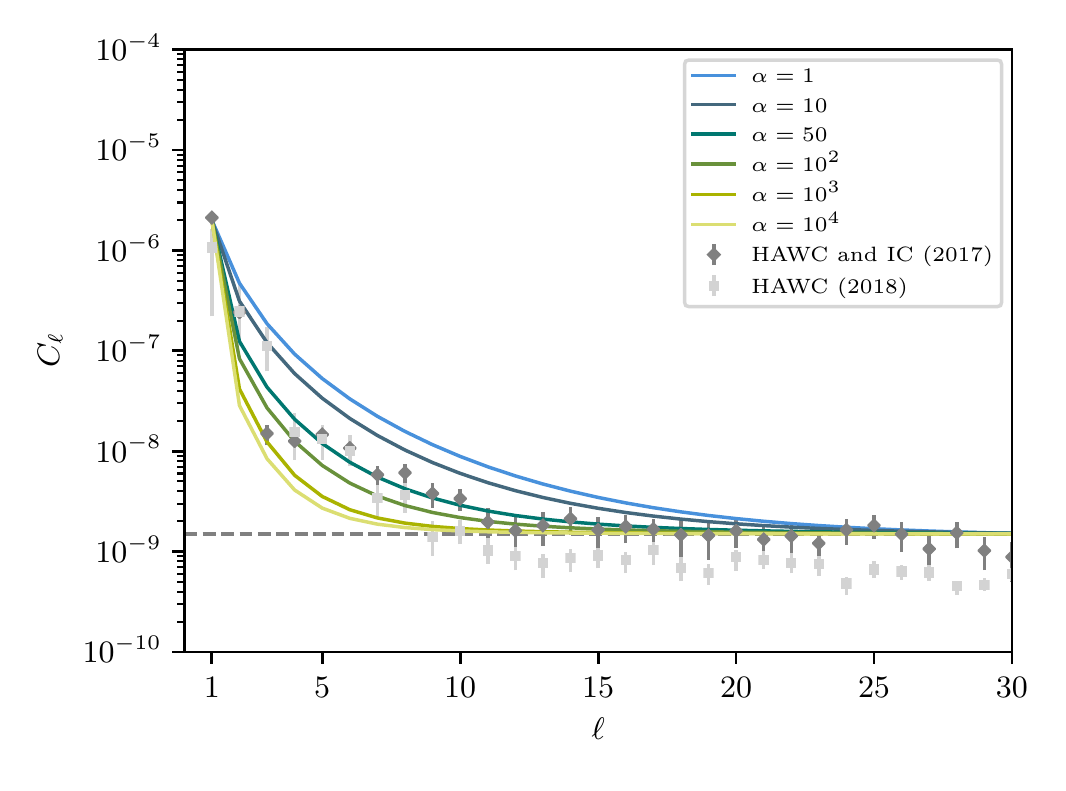}
\caption{The angular power spectrum computed in quasi-linear theory for different values of $\kin{}  \Delta T$, without (upper panel) and with (lower panel) adding the noise contribution. For comparison, we also show the observations by HAWC~\cite{Abeysekara:2018qho} and the IceCube-HAWC combined data~\cite{TheHAWC:2017uyf} with the shot noise level estimated for the latter.}
\label{fig2}
\end{figure}

While the band-limited white-noise spectrum serves only as an approximation for the magnetic turbulence in our local environment, it is nevertheless instructive to compare our model predictions to the power spectrum observed by HAWC and IceCube~\cite{Abeysekara:2018qho,TheHAWC:2017uyf}. In Fig.~\ref{fig2} we show the steady-state angular power spectrum $C^{\text{stdy}}$ derived by numerically solving Eq.~\eqref{eqn:ODE_stdy}. In the upper panel, we have fixed the gradient to $K \left|{\nabla\overline{f}}/{\overline{f}}\right|^2 = 10^{-4}  \kout{}$ and show the result for a range of $\alpha = \kin{} \Delta T$. It can be seen that with increasing values of $\kin{} \Delta T$, the normalisation of the angular power spectrum is decreasing and the power spectrum tends to fall off much faster. We compare our model predictions to the angular power spectra inferred from HAWC data~\cite{Abeysekara:2018qho} and the combined IceCube-HAWC data~\cite{TheHAWC:2017uyf}. Note that we have not accounted for the shot noise necessarily present in the data or for cross talk between multipole moments due to IceCube's limited field of view, see Ref.~\cite{Ahlers:2016rox}. In the right panel of Fig.~\ref{fig2}, we do account for the effect of shot noise by adding a constant noise power of $\mathcal{N}\simeq1.5 \times 10^9$. This is reproducing the data from the combined analysis of HAWC and IceCube data~\cite{TheHAWC:2017uyf} which is dominated by shot noise above $\ell \gtrsim 10$. Overall, with $\alpha = \kin{}  \Delta T \simeq 50$, as suggested by the numerical simulations, see Sec.~\ref{sec:validation}, we find good agreement with the data, again for $K \left|{\nabla\overline{f}}/{\overline{f}}\right|^2 = 10^{-4}  \kout{}$.

\section{Summary and Conclusion}
\label{sec:summary}

We have presented a computation of the angular power spectrum of CR small-scale anisotropies, based on the idea that the small-scale anisotropies are a consequence of cosmic ray streaming in the local configuration of the turbulent magnetic field. This model is based on a formal solution to the evolution equations for pairs of CR particles, and expressed as a steady-state solution of the ensemble-averaged products of their phase-space densities. We have evaluated this solution in a perturbative approach which can be represented by a series of diagrams. Considering only the contributions from the lowest order terms, we have formulated an ordinary differential equation for the angular power spectrum and solved for its steady state.

We have assumed throughout the absence of a regular magnetic field such that the unperturbed trajectories are straight lines. In order to formulate the ordinary differential equation, we also needed to adopt a band-limited white noise power spectrum for the turbulent magnetic field. This introduced two free parameters, the inverse of the smallest turbulent scale, $\kin{}$, and the smallest time-interval considered, $\Delta T$, but the steady-state angular power spectrum only depends on their combination $\alpha = \kin{}  \Delta T$. By comparing to numerical test particle simulations we have found a value of $\alpha \simeq 50$ to be appropriate. With this value, we find good agreement between our model predictions and the measurements by HAWC and IceCube.

The most obvious limitations of the present model are the unrealistic power spectrum that needs to be adopted and the dependence on the parameter $\alpha$ that needed to be fixed with the help of numerical simulations. We are convinced, however, that both are artefacts introduced by the fact that we treat unperturbed trajectories as straight lines. While we have motivated this by the assumed absence of a regular magnetic field, it is true that even in that case particles will experience an average regular field, set by the largest scales on which there is significant power.

In the future, it would therefore be desirable to consider a regular background magnetic field and unperturbed helical trajectories. This will also introduce resonance effects between the particles' gyroradii and the turbulent wavelengths which are also absent due to the assumed straight-line trajectories. We stress that such resonances and the form or the turbulent power spectrum are ultimately responsible for the energy-dependencies of the pitch-angle scattering rate and of the spatial diffusion coefficients. Observationally, this would broaden the range of predictions of our model. Given that the HAWC collaboration has already started presenting angular power spectra for different energy bins~\cite{Abeysekara:2018qho}, this avenue seems most promising.

\section*{Acknowledgments}

This work was supported by Danmarks Grundforskningsfond under grant no.\ 1041811001. MA also acknowledges support from \textsc{Villum Fonden} (project no.~18994).

\bibliographystyle{jhep}
\bibliography{qualinani}

\end{document}